\documentclass{acm_proc_article-sp}
\usepackage{amssymb}
\usepackage{algorithm}
\usepackage{algorithmic}

\usepackage{graphicx}
\usepackage{color}
\usepackage{pstricks,pst-node,pst-tree}
\begin{document}

\title{{WaterFowl, a Compact, Self-indexed RDF Store with Inference-enabled Dictionaries}
}

\numberofauthors{4} 
%
\author{
%
%
\alignauthor
Olivier Cur\'e\\
       \affaddr{Universit\'e Paris-Est}\\
       \affaddr{LIGM UMR CNRS 8049, France}\\
       \email{ocure@univ-mlv.fr}
\alignauthor
Guillaume Blin\\
       \affaddr{Universit\'e Paris-Est}\\
       \affaddr{LIGM UMR CNRS 8049, France}\\
       \email{gblin@univ-mlv.fr}
\alignauthor 
Dominique Revuz\\
       \affaddr{Universit\'e Paris-Est}\\
       \affaddr{LIGM UMR CNRS 8049, France}\\
       \email{drevuz@univ-mlv.fr}
\and
\alignauthor 
David Faye\\
       \affaddr{Universit\'e Gaston Berger de Saint-Louis}\\
       \affaddr{LANI, S\'en\'egal}\\
       \email{dfaye@igm.univ-mlv.fr}
}

\maketitle
\begin{abstract}
In this paper, we present a novel approach -- called WaterFowl -- for the storage of RDF triples that addresses some key issues in the contexts of big data and the Semantic Web. 
The architecture of our prototype, largely based on the use of succinct data structures, enables the representation of triples in a self-indexed, compact manner without requiring decompression at query answering time. 
Moreover, it is adapted to efficiently support RDF and RDFS entailment regimes thanks to an optimized encoding of ontology concepts and properties that does not require a complete inference materialization or extensive query rewriting algorithms. 
This approach implies to make a distinction between the terminological and the assertional components of the knowledge base early in the process
of data preparation, $i.e.$, preprocessing the data before storing it in our structures.
The paper describes the complete architecture of this system and presents some preliminary results obtained from evaluations conducted on our first prototype. 
\end{abstract}

%
%

\section{Introduction}

The emergence of big data imposes to face important data management issues: the most predominant ones being scalability, distribution, fault tolerance and low latency query answering.
The current trends in handling large volumes of information focus on parallel processing with MapReduce \cite{DBLP:conf/osdi/DeanG04} inspired frameworks. 
We consider that, for at least cost efficiency reasons, this approach may soon not be satisfactory anymore and should be combined with local data compression, $i.e.$, on each machine of a cluster. 
Hence, one will get the most out of a data center by distributing a data set over a cluster of machines and by compressing each partition in a clever way.

RDF (Resource Description Framework), a data model proposed by the W3C to represent metadata about Web resources, is totally concerned with this phenomenon.
This is partly due to the production of an increasing number of voluminous data sets, $e.g.$, 32 billion triples in the Linked Open Data cloud in March 2013. 
In this standard, a triple is made up of a subject, a property and an object and is generally represented as a graph. 
To foster interoperability among applications manipulating RDF data, vocabularies such as RDFS (RDF Schema) and OWL (Web Ontology Language) have been defined in the context of the W3C's Semantic Web Activity.
They support further means to describe the structure and semantics of RDF graphs and are themselves expressed as RDF triples.
When considered together, RDF data and its vocabulary represent a knowledge base which presents the main advantage of consistently managing the data and metadata within the same data model.
In the context of a Semantic Web knowledge base, handling inferences within query processing adds to the list of previously cited database management issues. 
Moreover, partitioning graph oriented data (a process needed in data distribution) is known to be a hard problem 
which may be more involved than sharding a relational database. 
Several solutions -- such as Virtuoso\footnote{http://virtuoso.openlinksw.com/}, 4Store\footnote{http://4store.org/} and OWLIM\footnote{http://www.ontotext.com/owlim} -- have
already tackled the problem of distributing triples over a cluster of machines.
But they do not consider high compression representations of the triples and we can consider that
developing such systems is still a challenging and crucial open problem. 

In this paper, we design a new architecture for RDF database systems that addresses compression and inference-enabled query answering and evaluate it using a proof of concept prototype. 
This framework will serve as the cornerstone for upcoming features that will include
data partitioning and supporting data updates.
The foundation of our system consists of a high compression, self-indexed storage structure supporting data retrieving decompression-free operations. 
By self-indexed, we mean that one can seek and retrieve any portion of the data without accessing the original data itself. 
Succinct Data Structures (henceforth SDS) provide such properties and are extensively used in our architecture via wavelet trees. 
The high rate compression obtained from SDS enables the system to keep all the data in-memory and limits latencies associated with Input/Output operations (see Section \ref{experiments}) via an efficient serialization/deserialization solution. 
Based on a preliminary work of Fern\'andez \textit{et al.} called HDT (Header Dictionary Triples) \cite{DBLP:conf/semweb/FernandezMG10} -- considered as a first attempt in this direction -- we 
propose to push its inner concept further to its logical conclusion by relying exclusively on bit maps and wavelet trees at all levels of our architecture (see Section \ref{system_description}). 
Moreover, the used data structures motivate the design of an original query processing solution that integrates efficient optimization and RDFS inferences which were not considered in \cite{DBLP:conf/semweb/FernandezMG10} nor in \cite{DBLP:conf/esws/Martinez-PrietoGF12}. 
The basic idea is to use an encoding of the data that will capture the subsumption relationships of both concepts and properties. 
Therefore, the encoded data will enclose -- without extra cost -- both raw data and ontology hierarchies. 
To efficiently use this encoding, the system will need to adapt standard \emph{rank} and \emph{select} wavelet tree operations into
ones that consider prefix of binary encoded identifiers \cite{DBLP:conf/pods/GrossiO12}.
This solution will spare the use of an expensive query rewriting approach \cite{DBLP:conf/semweb/Perez-UrbinaHM09}  or complete inference materialization (via a forward-chaining approach) when 
requesting a given ontology element, $i.e.$, concept or property, and all its sub-elements.
In order to complete RDFS entailment regime, we address \emph{rdfs:domain} and \emph{rdfs:range} through a minimalist materialization of subject, respectively object, \emph{rdf:type} properties.

This paper is organized as follows. In Section 2, we provide background knowledge on RDF and SPARQL as well as automata-based representations of a dictionary and SDS. 
Section 3 presents related work in the domains of RDF data management systems and query answering in the presence of inference throughout.
Section 4 details the main components of our architecture.  
Our proof of concept prototype is evaluated in Section \ref{experiments}.

\section{Background}
\subsection{RDF and SPARQL}
Assuming disjoint infinite sets U (RDF URI references), B (blank nodes) and L (literals), a triple (s,p,o) $\in$ (U $\cup$ B) x U x (U $\cup$ B $\cup$ L) is called an RDF triple with s, p and o 
respectively being the subject, predicate and object. 
Intuitively, a predicate denotes the relationship between subject and object.
This can be represented as an oriented labeled graph where the nodes are the subjects and objects and the labeled directed edges are the predicates.

SPARQL\footnote{http://www.w3.org/TR/rdf-sparql-query/} is the official W3C recommendation for querying RDF data and is generally composed of triple patterns called Basic Graph Pattern (BGP).
The computation of a query answer set generally involves graph pattern matching.
In order to define BGP, we now also assume that V is an infinite set of variables and that it is disjoint with U, B and L. 
We can recursively define a SPARQL graph pattern as follows: 
(i) a triple $gp \in$ (U $\cup$ V) x (U $\cup$ V) x (U $\cup$ V $\cup$ L) is a SPARQL graph pattern, 
(ii) if $gp_1$ and $gp_2$ are graph patterns, then ($gp_1 . gp_2$) represents a group of graph patterns that must all match, $i.e.$, the dot 
operator corresponds to a conjunction, ($gp_1$ \texttt{OPTIONAL} $gp_2$) where $gp_2$ is a set of patterns that may extend 
the solution induced by $gp_1$, $.i.e.$, similar to an outer join of relational algebra, and ($gp_1$ \texttt{UNION} $gp_2$), denoting pattern alternatives, are graph patterns and 
(iii) if $gp$ is a graph pattern and C is a built-in condition then the expression ($gp$  \texttt{FILTER} C) is a graph pattern that enables the restriction of the solutions of a graph pattern match according 
to the expression C.
The SPARQL syntax follows the \texttt{SELECT-FROM-WHERE} approach of SQL queries. The \texttt{SELECT} clause specifies the variables appearing in the result set of the query. 

In \cite{harris_sparql11_2013}, extensions to SPARQL semantics, called entailment regimes, are presented. 
In this work, we address RDF and RDFS entailment regimes in the context of skolemization as presented in  \cite{harris_sparql11_2013}, $i.e.$, a syntactic transformation that replaces  blank nodes  by 'new' names. 
Most of the inferences we are considering are related to entailment rules proposed in \cite{hayes_rdf_2004}. 
An RDF Schema describes semantic constraints between classes and properties used in an RDF graph.
These description are defined in terms of RDFS built-in properties that support subclass and subproperty relationships, resp. \emph{rdfs:subClassOf} and \emph{rdfs:subPropertyOf},
class typing to both the domain and the range of a property, resp. \emph{rdfs:domain} and \emph{rdfs:range}.
In the context of query answering, these rules are useful to check satisfiability and rewrite queries. 
In our system, they are implemented through the use of adapted encodings and data structures which are directly motivated by SDS.

\subsection{Automata-based dictionary of non-concept property elements}
\label{automata}
The purpose of the dictionary is to support the encoding of URIs, blank nodes and literals encountered in the subject and object positions, URIs in properties of instance triples. 
Remember that we do not consider the ontology (terminological box) and the instances (assertional box) are the same level.
In Section \ref{onto} we will present the encoding approach used for the ontology. In this section, we consider the structures of the elements found in the instances.
This encoding takes the form of a key/value pair where the key is a unique integer and the value is either an URI, a blank node or a literal.
Our system requires two-way access to this structure: from keys to values (\textit{i.e.}, to translate the result set of a query) and from values to keys (\textit{i.e.}, 
to encode the triples, to translate the SPARQL queries from URIs and strings to identifiers as well as to handle SPARQL \texttt{FILTER} clauses)
Note that this dictionary approach, an initial compression strategy, is frequently encountered in triple store solutions but usually, provided without any implementation details.
The dictionary implementation adopted in WaterFowl is, depending on the data set, based on either an automaton or a trie. 
Automata are very efficient data structures for representing natural language lexicons \cite{revuz91}, with efficient time and space complexities. 
In data sets, the strings associated with subjects, predicates and objects are of two kinds\,: a set of similar strings (sharing common prefixes - \textit{e.g.}, namespaces) and a set of singletons. 
Automaton can be stored in memory and act as a reversible mimimal perfect hash function on the set of strings. 
Considering the set of singletons, automata do not provide a compression gain. 
But, one may store the singletons into a flat file on disk and use a trie in order to bind a given prefix to the position of the corresponding word in the file. 
With the right file system implementation, this yields a unique file access per string. 
These techniques of minimal perfect hashing with DAWG (Directed Acyclic Word Graphs) or Tries enable us to encode any subject, predicate and object as an integer which will be used in the 
compressed version of the set of the triples. 
Futhermore, independently to where they appear (as subject or object) we can use a common encoding due to our layered architecture (see Section 4) -- leading to a gain of space. 
All related implementations are realized using the C++ template library ASTL\footnote{http://astl.sourceforge.net}. 

\subsection{Succinct Data Structures}
\label{sds}
The family of SDS uses a compression rate close to theoretical optimum, but simultaneously allowing efficient decom\-pression-free query operations on the compressed data.
This property is obtained using a small amount (o(Z) bits where Z corresponds to the theoretical optimum) of extra bits to store extra information. 
Initially introduced by Jacobson \cite{DBLP:conf/focs/Jacobson89} when considering bit vectors, the concept is nowadays extended to wider alphabets.

Bit vectors (aka bit maps) are useful to represent data while minimizing its memory footprint. 
In its classical shape, a bit vector allows, in constant time, to access and modify a value of the vector. 
Munro \cite{DBLP:conf/fsttcs/Munro96} designed an asymptotic optimal version where, in constant time, one can (i) count the number of 1 (or 0) appearing in the first x elements of a 
bit vector (denoted \emph{$rank_b(x)$} with b $\in$ \{0,1\}), (ii) find the position of the x$^{th}$ occurrence of a bit (denoted \emph{$select_b(x)$}, b $\in$ \{0,1\}) and 
(iii) retrieve the bit at position x (denoted \emph{$access(x)$}). 
In the remaining of this paper, we do not precise the bit $b$ anymore and simply write \emph{rank} and \emph{select}.

Naturally, these operations on bit vectors would be of great interest for a wider alphabet. 
The original solution was provided by Grossi \textit{et al.} \cite{DBLP:conf/soda/GrossiGV03} and roughly consists in using a balanced binary tree -- so-called wavelet tree. 
The alphabet is splitted into two equal parts. One attributes a 0 to each character of the first part and a 1 to the others. 
The original sequence is written, at the root of the tree, using this encoding. 
The process is repeated, in the left subtree, for the subsequence of the original sequence only using characters of the first part of the alphabet and, in the right subtree, for the second part. 
The process iterates until ending up on singleton alphabet. Roughly, one has provided an encoding of each character of the alphabet. 
Using \emph{rank} and \emph{select} operations on the bit vectors stored in the nodes of the tree, one is able to compute \emph{rank} and \emph{select} operations on the original sequence in O(log $\vert$alphabet$\vert$) by deep 
traversals of the tree. These operations can be easily adapted to only traverse until a given depth -- refered as \emph{rank\_prefix} and \emph{select\_prefix} operations (that will be of great interest for us along with 
our encoding of ontology concepts and properties). Wavelet trees have been well studied since then and both space and time efficient implementations are now available ($e.g.$, pointer-free wavelet 
tree and wavelet matrix of the libcds library\footnote{https://code.google.com/p/libcds/}).

\section{Related work}
In this section, we consider related work in the fields of indexing RDF data sets, query processing in the presence of inferences and 
query optimizations. 

Although the first RDF stores appeared in 2002, $e.g.$, \cite{DBLP:conf/semweb/BroekstraKH02}, this research field became really active in 2007 starting
with the publication of \cite{DBLP:conf/vldb/AbadiMMH07}. 
Before that paper, most systems where storing their triples in a relational database management system as the backend storage, using different approaches, $e.g.$, 
triple table ($i.e.$, a single table with 3 columns for s,p and o) or different variants such as clustered property table or property-class table.
Abadi \textit{et al}'s paper motivated the development of systems which were not using relational database management systems as a storage layer and were considering indexes with more attention than previous solutions. 
Hence, solutions such as Hexastore \cite{DBLP:journals/pvldb/WeissKB08} and RDF-3X \cite{DBLP:journals/vldb/NeumannW10} were designed using  multiple indexes, respectively 6 and 15, which had a direct 
impact on the performance of query answering but also on the memory footprint of databases.
Matrix Bit loaded \cite{DBLP:conf/www/AtreCZH10} is another multiple indexes solution which stores its data into bit matrices. 
Compared to these systems, our approach proposes a single structure that enables indexed access on the three components of the triples.

Our approach is inspired from the HDT \cite{DBLP:conf/semweb/FernandezMG10} solution which mainly focuses on data exchange (and thus on data compression). 
Its former motivation was to support the exchange of large data sets highly compressed using SDS.
Later, \cite{DBLP:conf/esws/Martinez-PrietoGF12} presented HDT FoQ, an extension of the structure of HDT that enables some simple data retrieving operations.
Nevertheless, this last contribution was not allowing any form of reasoning nor was providing methods the query the data sets.
In fact, WaterFowl brings the HDT FoQ approach further to its logical conclusion by using a pair of wavelet trees in the object layer (HDT FoQ uses an adjacency list for this layer) 
and by integrating a complete query processing solution with complete RDFS reasoning (\textit{i.e.}, handling any inference using RDFS expressiveness). 
This is made possible by an adaptation of both the dictionary and the triple structures. 
Note that this adaptation enables to retain the nice compression properties of HDT FoQ (see Section 5).

\begin{figure*}
\centering
\includegraphics[scale=0.9]{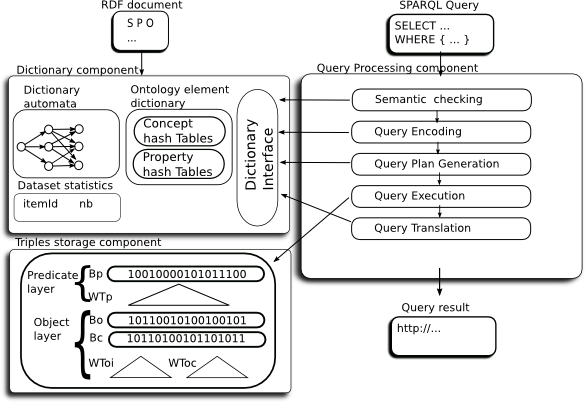}
\caption{WaterFowl's architecture}
\label{archi}
\end{figure*} 

Concerning query processing in the presence of inferences, several approaches have been proposed. 
Among them, the materialization of all inferences within the data storage solution is a popular one, which is generally performed using an off-line forward chaining approach. 
This avoids query formulation at run-time but is associated with an expansion of the memory footprint. 
Sesame\footnote{http://www.openrdf.org/} is a commercial system adopting inference materialization. 
Another approach consists in performing query rewriting at run time. It guarantees a light memory footprint but it is associated with the possible generation of an exponential number of queries. 
Presto \cite{DBLP:conf/kr/RosatiA10} and Requiem \cite{DBLP:conf/semweb/Perez-UrbinaHM09} are system adopting this approach with different algorithms.
By adopting a rewriting approach into non recursive datalog, Presto achieves to perform this operation in non exponential time.
The technique proposed in \cite{DBLP:conf/kr/GottlobS12} produces worst-case polynomial rewritings but the complex structure it is based on makes its evaluation complex to perform.

The encoding of ontology elements, $i.e.$, concepts and properties, used in our system is related to a third approach which consists in 
encoding elements in a clever way that retains the subsumption hierarchy.
This is the approach presented in \cite{DBLP:conf/kr/Rodriguez-MuroC12} and implemented in the Quest system (a relational database management system). 
The work of Rodriguez-Muro $et\,al.$ \cite{DBLP:conf/kr/Rodriguez-MuroC12}  relies on integer identifiers modeling the subsumption relationships which are being used to rewrite SQL queries ranging over identifiers intervals, $i.e.$, 
specifying boundaries over indexed fields in the WHERE clause of a SQL query. 
In comparison, our work tackles the encoding at the bit level and focuses on the sharing of common prefixes in the encoding of the identifiers (see Section 4.1). 
This approach allows us to rewrite the queries in terms of  \emph{rank\_prefix} and \emph{select\_prefix} operations, $i.e.$, searching for a pattern corresponding to some of the most significant bits of a 
concept or property identifier. Furthermore, it allows high rate compression and does not require extra specific indexing processes.

Finally, our solution focuses on query processing of SPARQL queries. 
It aims to minimize the memory footprint required during query execution and to perform optimizations in terms of SDS operations complexities: \emph{access}, \emph{rank}, \emph{rank\_prefix}, \emph{select} and \emph{select\_prefix}. 
Adapting some of the heuristics presented in \cite{DBLP:conf/edbt/TsialiamanisSFCB12}, optimization approaches of \cite{DBLP:conf/sigmod/NeumannW09} and \cite{DBLP:conf/www/StockerSBKR08}, the system optimizes execution of SDS operations. 
The ordering of basic graph patterns execution also takes into account simple statistics computed when generating the dictionaries (see Section \ref{query}).
BigOWLIM\footnote{http://www.ontotext.com/owlim/} is, like most existing RDF database systems, $e.g.$, RDF-3X and Jena TDB\footnote{http://jena.apache.org/documentation/tdb/}, 
taking benefits of data statistics to organize the order of BGP and thus optimize queries.

\section{System description}\label{system_description}
We now describe the main components (Figure \ref{archi}) of WaterFowl's current architecture.
It is composed of three components: dictionary, triple storage and query processing.
The functionalities implemented in each of these components are detailed in the following subsections.

\subsection{Dictionary component}
\label{onto}
The aim of dictionaries is to reply to the following expectations: (i) enable the transformation of the triples of SPARQL queries' WHERE clause, $i.e.$, transforming  URIs and literals to their corresponding integer
identifiers, (ii) allow the transformation of integer-encoded results obtained from the query processing component into URIs, blank nodes and literals and 
(iii) support various inference-related operations such as a form of query transformation and semantic checking. 
Note that objectives (i) and (ii) are shared with our automata-based dictionary approach.  This component contains several data structures which are organizing the storage and the access of different dictionaries.
An automata-based dictionary stores all non ontology and non vocabulary elements, $i.e.$, it does not store any of the concepts and properties of the ontology nor entries of the RDFS vocabulary ($e.g.$, \emph{rdf:type}).
Section \ref{automata} already provided sufficient details and motivations about this structure and we will not concentrate on it any further.
 
As presented in \cite{DBLP:conf/sigmod/NeumannW09}, the kind of histograms that support
query optimization in a relational context cannot be transposed to RDF graphs due
to the prohibitive size that would be needed to store them and the amount of time needed 
to compute them.
Hence, a restricted amount of data set statistics are stored in this component.
It amounts to storing the total number of subjects, predicates and objects in the data set as well as statistics on triples distribution,
$i.e.$, the number of occurences of distinct subjects, predicates and objects.
These statistics mainly serve to help in finding the most cost-efficient physical plan of a given query. We will provide more details in Section \ref{query}. The dictionary interface supports the communication between the query processing and the dictionary components, $e.g.$, use data set statistics, encode a query's triple patterns and decode an answer set.

\begin{figure*}
\centering
\includegraphics[scale=0.8]{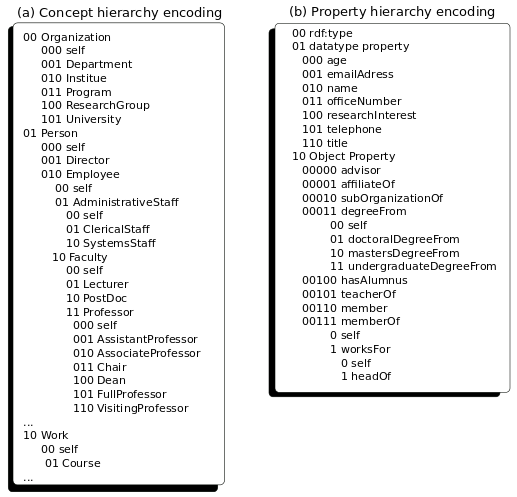}
\caption{Encoding for an extract of LUBM's ontology hierarchies}
\label{encodings}
\end{figure*} 

In the remaining of this section, we concentrate on the ontology element dictionary, $i.e.$, concepts and properties (one for each). 
The generation of the ontology element dictionaries is performed off-line and we currently do not consider ontology updates. 

The ontology encoding is characterized by integer identifiers attributed to each ontology element entry.
These integer values are possibly shared with entries of our other dictionaries but this is not an issue since we contextualize them. 
That is, we know that each value appearing in the second position of a triple or of a SPARQL BGP is necessarily a property. 

Similarly for concept identifiers, we know that in the data set their appearances as an object are associated with an \emph{rdf:type} property. 
Since our method to handle SPARQL BGPs is based on navigating through our 
two-layered structure, we always get the information required to consider the context.
This identifier sharing characteristic among our different dictionaries opens up the encoding of very large set of identifiers, regardless of the structure of concept and property hierarchies. 
We will see that the distribution of identifiers generated for the ontology dictionaries is qualified by a possibly high sparsity. 
Hence, enabling an encoding over a very large sets of identifiers ensures to support very large data sets and ontologies.

Our encoding methodology is directly motivated by the structure we are using in our two-layered data structure (detailed in Section \ref{structure}). 
The overall objective is to encode the data itself and the ontology hierarchies (that is the subsumption relations) in a compact way. 
To do so, the encoding will include in its definition the information of subsumption. 

Prior to encoding, we are using a Description Logic reasoner, $e.g.$, Pellet\footnote{http://clarkparsia.com/pellet/} or HermiT\footnote{http://hermit-reasoner.com/}, 
to perform of classification of concepts. Note that this approach enables to consider 
ontologies more expressive than RDFS, $e.g.$, OWL.
Then, we navigate in a depth-first search manner through this classication.
This enables to compute the representation of all concepts such that any pair of concepts sharing a common ancestor in the concept hierarchy will share a common prefix in their representation (corresponding to this common ancestor).

To do so, starting from the \emph{owl:Thing} and an empty prefix, we compute the number of direct subconcepts of \emph{owl:Thing}. We encode each of these last with a minimum number of bits. This encoding will be a common prefix to any concept belonging to the hierarchy depending on the subsumption relation. Figure \ref{encodings}(a) represents an extract of LUBM's ontology. 
It emphasizes that \texttt{owl:Thing}'s direct subsumption hierarchy is encoded on 2 bits and that any subconcept of \emph{Organization} (resp. \emph{Person} and \emph{Work}) is encoded 
with prefix $00$ (resp. $01$ and $10$). 

We will now act in a similar way for each direct subconcepts of \emph{owl:Thing}. The only difference will be that we will assume that any concept (except \emph{owl:Thing}) has a direct subconcept named \emph{self}. This assumption is needed to differentiate, in query processing, a query targeting a given concept (referred as \emph{self}) or its set of subconcepts.  
For ease of treatment, we will always attribute the $0$ value to \emph{self}. Hence the encoding associated to \emph{self} will correspond to a given concept (as if it was a subconcept of itself) while the identifier of the concept corresponds to its set of subconcepts.
For example, querying any concept encoded with the prefix $00$ will correspond to seeking for any kind of \emph{Organization} while querying any concept encoded with the prefix $00\;000$ will seek specifically for \emph{Organization} excluding its subconcepts. Indeed, the prefix $00\;000$ excludes \emph{Department} which is encoded by the prefix $00\;001$ while the prefix $00$ includes all kind of \emph{Organization}. By recursively processing the hierarchy of concepts, one will end up with a prefix encoding (as illustrated in Figure \ref{encodings}). 

This \emph{self} mechanism is not required for \emph{owl:Thing} since it is handled natively within our framework. 
Provided with this encoding one can easily query any entry regarding a given concept and its subconcepts by the use of \emph{rank\_prefix} and \emph{select\_prefix} operations.


Considering the properties, we first distinguish between the \emph{rdf:type}, datatype and object properties encountered in the data sets and assign specific prefixes of $2$ bits to each of them. For ease, we will always attribute the prefix $00$ to \emph{rdf:type}. For both the sets of object and datatype properties, we apply a similar process as for the concepts in order to achieve a prefix encoding. Figure \ref{encodings}(b) displays the property encodings for an extract of the LUBM's ontology.

The corresponding encodings are stored in two types of hash tables: (i) one with an identifier as key and URI as value, denoted $H_1$, and (ii) one with URI as key and a tuple consisting of (a) an identifier, 
(b) the number of bits required to encode the direct sub-elements of this element and (c) some additional parameters such as number of occurrences
, denoted $H_2$.
This additional informations are necessary to allow for the completeness of the RDFS entailment regime and to detect unsatisfiable queries, $e.g.$, 
when a SPARQL variable is bounded to a concept $C$ that is not instantiated in the data set, which may require inferences, $i.e.$, modifying the query such that the variable ranges over the subconcepts of $C$. 
It is also useful for reordering graph patterns to minimize the memory footprint of the executed query. 
For example, considering data sets generated from the LUBM, there is no instance for the \emph{Professor} concept and LUBM's query \#4 is unsatisfiable. 
Nevertheless, this query returns some results if the system seeks for all subconcepts of \emph{Professor}. 

Our approach is, by far, adapted to tree-like hierarchies. Nevertheless, we can support multiple inheritance of ontology entities in several ways. First of all, in order to capture all the knowledge, one would have to use different prefixes for the same ontology entity. For example, let us consider a concept $A$ having $X$ and $Y$ as super-concepts respectively identified by the prefixes $00$ and $01$. One solution would be to materialize two copies of each fact inducing $A$ using the two encodings induced by the inheritance of $X$ and $Y$ -- let say using the prefixes $00\;01$ and $01\;10$. In order to retrieve any knowledge, one would also need to duplicate the information concerning subconcepts of $A$. We claim that this approach is not very efficient since it will dramatically increase the size of the data set and the query answering time due to unnecessary unions to cover all the occurrences.

The solution we adopted relies on  providing a single prefix to any concept -- even the one with multiple super-concepts. Arbitrarily, we decide to assign the prefix corresponding to the first super-concept encountered in the data. Hence, all occurrences of a concept in the data set will share a single common prefix. There will be no  expansion of the data set. In order to be able to derive all the knowledge induced by the multiple inheritance we, moreover, use an additional data structure that provides possible encodings of any concept. 

Considering our previous example, we store in this data structure that concept $A$, which appears in the data as $01\;10$, can also be seen as encoded as $00\;01$. Our solution, will thus use some query rewriting techniques in order to retrieve all informations induced by the multiple inheritance. For example, if one wants to retrieve all information regarding any sub-concept of $X$, this request should require any concept encoded using the prefix $00$. 
In the data, since, arbitrarily, the encoding of $C$ is $01\;10$, in order to retrieve any information of $C$ or of one of its sub-concepts, one will ask for the union of any concept with prefix $00$ or $01\;10$ since in the equivalence data structure, $00\;01$ (which is a candidate) is equivalent to $01\;10$. This is simply performed with the addition of \texttt{UNION} clause in the SPARQL query. Even if this approach also has a drawback (possibly heavy query rewriting), one only needs to efficiently know which subsumption relation are not directly expressed in the data and to store multiple inheritance for the direct common sub-concept only (which clearly are rare). On the whole, this solution seems more acceptable for our purpose than heavy materialization.




\subsection{Triples storage component}
\label{structure}
Once the dictionaries have been defined
, the triples can be encoded 
 in a structure that makes an intensive use of SDS. 
 To illustrate the structure, we will encode the following simple RDF triples.

\vspace*{-0.5cm}
\begin{verbatim}
Uni0.edu rdf:type ub:University
Uni0.edu ub:name "University0"
Dpt0.Uni0.edu rdf:type ub:Department
Dpt0.Uni0.edu ub:name "Department0"
Dpt0.Uni0.edu ub:subOrganizationOf Uni0.edu
Dpt0.Uni0.edu/AP0 rdf:type ub:AssociateProfessor
Dpt0.Uni0.edu/AP0 ub:name "Cure"
Dpt0.Uni0.edu/AP0 ub:teacherOf Dpt0.Uni0.edu/C15
Dpt0.Uni0.edu/AP0 ub:teacherOf Dpt0.Uni0.edu/C16
Dpt0.Uni0.edu/AP0 ub:worksFor Dpt0.Uni0.edu
Dpt0.Uni0.edu/C15 rdf:type ub:Course
Dpt0.Uni0.edu/C15 ub:name "Course15"
Dpt0.Uni0.edu/C16 rdf:type ub:Course
\end{verbatim}

\begin{figure*}[t]
\centering
\includegraphics[scale=0.7]{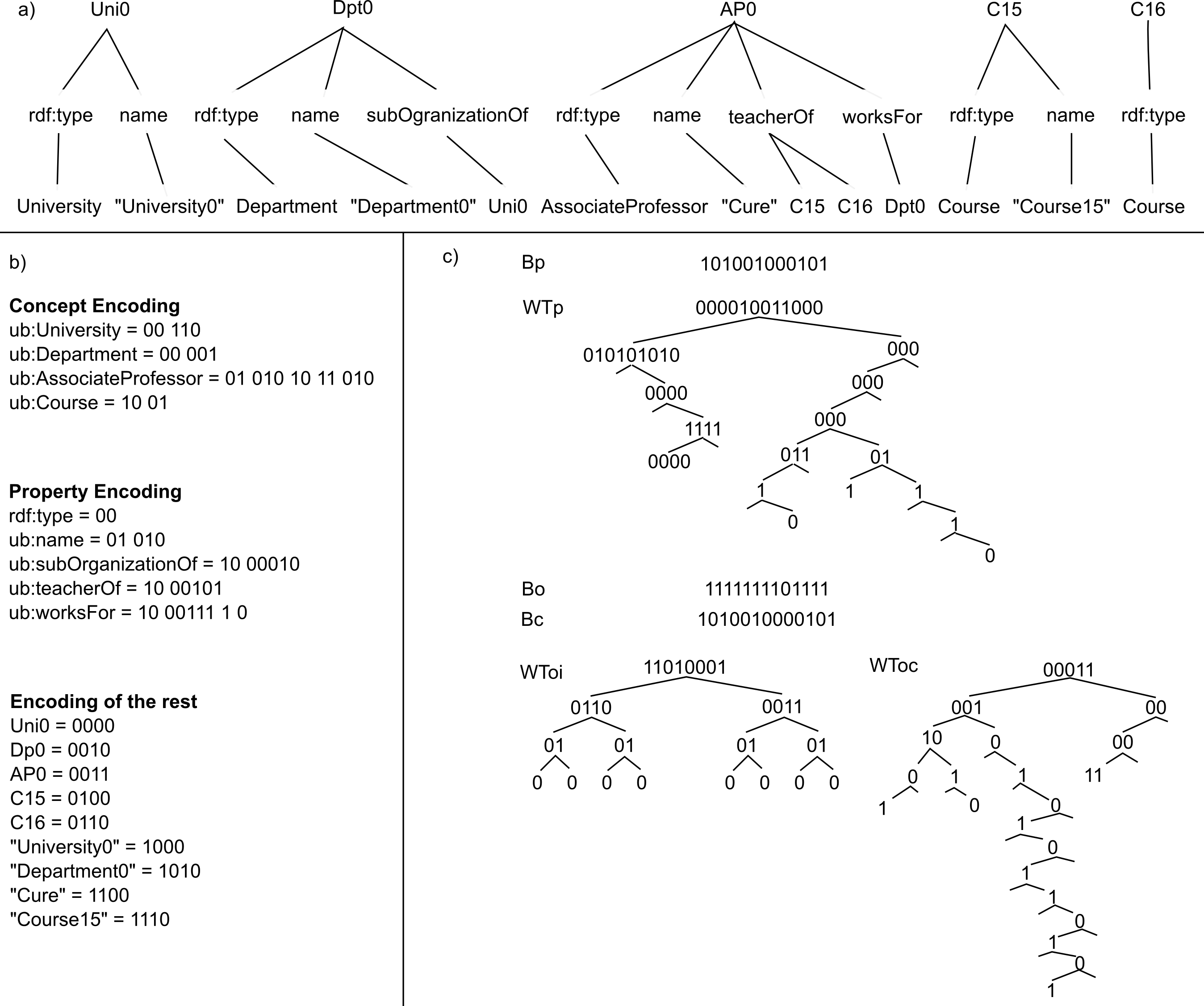}
\caption{Two-layer structure. For ease of presentation, URIs have been shorten. a) Tree-like representation of some RDF triples. b) Encodings. c) Corresponding storage.}
\label{triples}
\end{figure*}

\vspace*{-0.5cm}
To do so, the triples are ordered by subjects, predicates and then objects. 
To simplify the understanding of our storage structure, we represent the ordered set of triples as an ordered forest (Figure \ref{triples}(a)) which will serve
to demonstrate the creation of our two-layer structure where each layer is composed of bitmaps and wavelet trees.

The first layer encodes the relation between the subjects and the predicates; that is the edges between the root of each tree and its children. 
The bitmap $B_p$ is defined as follows. For each root of the trees in (Figure \ref{triples}(a)) -- that is each subject -- the leftmost child is encoded as a $1$, and the others as a $0$. 
On the whole, $B_p$ contains as many $1$'s as subjects in the data set and is of length equal to the number of predicates in the data set. In Figure \ref{triples}(c), one obtains $101001000101$ since there are $5$ subjects with the last subject having $1$ predicate, the first and fourth subjects having $2$ predicates, the second one having $3$ while the third one having $4$. The wavelet tree $WT_p$ encodes the  sequence of predicates obtained from a pre-order traversal in the forest ($e.g.$, second row in Figure \ref{triples}(a)). The construction of the wavelet tree follows the algorithm presented in Section \ref{sds}.

Unlike the first layer, the second one has two bitmaps and two wavelet trees. $B_o$ encodes  
the relation between the predicates and the objects; that is the edges between the leaves and their parents in the tree representation. Whereas, the bitmap $B_c$ encodes the positions of ontology concepts in the sequence of objects obtained from a pre-order traversal in the forest ($e.g.$, third row in Figure \ref{triples}(a)).

The bitmap $B_o$ is defined as $B_p$ considering the forest obtained by removing the first layer of the tree representation (that is the subjects). In Figure \ref{triples}(a), one obtains $1111111101111$. The bitmap $B_c$ stores a $1$ at each position of an object which is a concept; a $0$ otherwise. This is processed using a predicate contextualization, $i.e.,$ in the data set whenever a \emph{rdf:type} appears, we know that the object corresponds to an ontology concept. In Figure \ref{triples}(a), considering that the predicate \emph{rdf:type} is encoding by $00$, one obtains $1010010000101$. Finally, the sequence of objects obtained from a pre-order traversal in the forest ($e.g.$, third row in Figure \ref{triples}(a)) is splitted into two disjoint subsequences; one for the concepts and one for the rest. Each of these sequences is encoded in a wavelet tree ($WT_{oc}$ and $WT_{oi}$ respectively). This architecture reduces sparsity 
of identifiers and enables the management of very large data sets and ontologies while allowing time and space efficiency.
 
\subsection{Query processing component}
\label{query}
The query processing component contains the modules displayed on the right part  of Figure \ref{archi}. 
It coincides with the classical modules found in standard relational database management systems. 
Nevertheless, these modules are adapted to optimize performances of query answering in the context of an RDF data model and SDS operations. 
Due to space limitations, this section details the aspects related to query processing involving inferences and only provides general information on the aspects not requiring any form of reasoning, 
$i.e.,$ we do not provide a complete presentation of our query optimization strategy which will be detailed in another paper.
In the remaining of this section, we will illustrate several aspects in the context of the LUBM \cite{DBLP:journals/ws/GuoPH05} ontology with the following SPARQL query (henceforth denoted $QR1$) 
which seeks for pairs of \emph{Professor/Department} satisfying the fact that the \emph{Professor} works for that \emph{Department}:

\vspace*{-0.5cm}
{\small
\begin{verbatim}
PREFIX rdf: <http://www.w3.org/1999/02/22-rdf-syntax-ns#>
PREFIX ub:<http://swat.cse.lehigh.edu/onto/univ-bench.owl#>
SELECT ?x ?y WHERE {?y rdf:type ub:Department. 
?x rdf:type ub:Professor. ?x ub:worksFor ?y.}
\end{verbatim}
}

\vspace*{-0.5cm}
A first step consists in the parsing of a SPARQL query and checking for its well-formedness. 
For each valid query, a semantic checking step is performed. 
It first involves to communicate with the dictionary component
to make sure that each element of a SPARQL graph pattern is present in the dictionaries. This is performed with both the automata based dictionary and the ontology element dictionaries through the use of a dictionary interface (Figure \ref{archi}) which receives a set of basic graph patterns. Given a triple context, the system seeks in the appropriate dictionary ($e.g.,$ search the object in the concept dictionary if the predicate is \emph{rdf:type}). The system detects two cases of unsatisfiability: (i) one of the graph pattern's element (excluding variables) is not present in any of the dictionaries, 
(ii) a graph pattern element has no occurrences in the data sets and, in the case of a concept or property has no instantiated sub-elements occurrences neither. 
Otherwise, the BGP is satisfiable and the module 
obtains identifiers and statistics associated to each non variable graph pattern element. Note that in the case of a concept or property element with sub-elements, it is the identifier associated to its \emph{self} counterpart that is returned.
In the case of $QR1$, the identifier and statistic associated to \emph{Professor} are respectively $01\;010\;10\;11\;000$, $i.e.,$  \emph{Professor}'s \emph{self} entry, and $0$ since LUBM's data sets do not instantiate directly this concept. This approach enables to detect unsatisfiable queries rapidly since it detects that the query's result set is empty without executing any other steps of the query processing component.
A query is considered unsatisfiable if one of its BGP is unsatisfiable otherwise, the whole query is satisfiable.

Another aspect of the \emph{semantic checking} component, and motivating its name,
depends on the expressiveness of the underlying ontology language, $i.e.$, RDFS or OWL.
The main idea is to consider the ontology axioms as constraints in  order to 
express if the SPARQL query is satisfiable or not.
This approach would enable to return an empty set to an unsatisfiable query without
requiring any query driven processing, $i.e.$, query translation, optimization and execution.
The topic of considering axioms of a Semantic Web ontology as data integrity constraints
has been the subject of many research works, $e.g.$, \cite{DBLP:conf/www/MotikHS07}.
The complexity comes from the open world and non unique name that one assumes in this context.
In the context of this paper, we are targeting RDFS entailment regime. Since, the RDFS 
language does not support negation nor concept/property disjunction, this semantic 
checking is not needed.
However the support of more expressive ontologies will benefit from this component, $e.g.$ RDFS++ or the EL and QL 
OWL2 profiles.

A satisfiable query is then encoded in terms of identifiers retrieved from the set of dictionaries. It results in a query containing integer-based graph patterns and variables. 
In this step, the statistics associated to concept and property ontology elements encountered in graph patterns of the query may imply some form of reasoning. For instance, consider that such a concept $C$ or property $P$ has no  instances, then since the query is satisfiable, it means that $C$ or $P$ has some sub-elements. Hence, some of its direct or indirect sub-elements may be instantiated and are expected in the result set of the query. The solution we are proposing is to replace the identifier of $C$ or $P$'s \emph{self} entry with $C$ or $P$'s own identifier, $i.e.,$ removing  \emph{self}'s local identifier in the query. In the context of $QR1$, it implies removing $000$, \emph{self}'s local identifier, from $01\;010\;10\;11\;000$ which yields to $01\;010\;10\;11$. It corresponds to the \emph{Professor} concept and is a common prefix 
to all its subconcepts.
This encoding is strongly linked to the notion of SDS prefixed operations and will only requires to partially navigate in the associated wavelet trees. 

One of the difficulty is to handle variable length prefixes in the query. To do so, a simple trick is to add a $1$ in front of any encoding. Considering our $01\;010\;10\;11$ identifier, we would then get $1\;01\;010\;10\;11$ ($i.e.$, $683$) where the first $1$ denotes the boundary of significant bits placed at its right. Querying any \emph{Employee} will correspond to $42$ ($i.e.$ $1\; 01\;010$).
Clearly, this trick allows us to provide both the prefix of interest and the number of significant bits which is related to the depth of the search in the wavelet trees.

So far, the presented solution has not opted for inference materialization and it means that database instances are not complete, $i.e.,$ they do not contain all implicit information.
Even in the case of the RDFS entailment regime, this pauses a problem with the \emph{rdfs:domain} and \emph{rdfs:range} properties.
To handle this incompleteness, we propose to infer and materialize the type of any concept according to the value of these two properties and the predicate the given concept is in relation with.
We have opted for this solution because it does not come at the cost of expanding our two layer structure, it does not imply any query rewriting and the memory footprint impact caused by this materialization is limited thanks to a non-naive and non-exhaustive approach of typing. Indeed, the idea is to infer the deepest concept -- say $D$ -- in the hierarchy that can match our target subject -- say $c$ -- and add a triple ($c$, \emph{rdf:type}, $D$) to the data set if it does not yet exists. Of course, any less expressive statement can also be removed from the data set.

Let us demonstrate this aspect with an example. In the LUBM ontology, the axioms $ \top \sqsubseteq \forall$ \emph{advisor}$^{-}.$\emph{Person} and $\top \sqsubseteq \forall$ \emph{advisor}$.$\emph{Professor} respectively defining that the \emph{advisor} property has the concept \emph{Person} as domain and \emph{Professor} as range. 
Now consider a data set where we have the following triples:\\
\emph{ex:smith ub:advisor ex:gblin.}\\
\emph{ex:gblin ub:worksFor ex:esipe.}\\
Moreover, we consider that no other triple explicitly types \emph{gblin} as a \emph{Professor} and provides any type to \emph{smith}. Then query $QR1$ would not return \emph{ex:gblin ex:esipe} in its answer set. 

Our approach is supported by performing a preprocessing step on the data set encoding step.
That is consider the triple $t$ with $(s,p,o)$ as respectively its subject, property and object.
We first search if $p$ has a known domain and/or range.
Let consider that it is the case for the domain (resp. range) then we search if $s$ (resp. $o$) was already typed in the data set. Note that this operation is not costly on the domain due to the ordering of our triple when encoding the data.
If the type $C$ of the triple element is a super concept of the domain (resp. range) then we do not have to add anything to our data set.
If this is not the case, $i.e.$, $C$ is a subconcept or in another concept branch of the hierarchy of the domain (or range) of $p$, then we add a new type triple stating that $s$ (resp. $o$) has a new 
type corresponding to the concept specified in the domain (resp. range) of  property $p$. In the case of our previous example, the following triples would be materialized.\\
\emph{ex:smith rdf:type Person.}\\
\emph{ex:gblin rdf:type Professor.}

In terms of computation of this preprocessing, dealing with \emph{rdfs:range}
is more involved since we cannot guarantee if an object will be typed and where it will
typed in an RDF data set.
This process needs to go throughout all its occurrences.
In our solution, we added another data structure that enables to support this triple addition efficiently.

A best effort query plan is then searched using a set of heuristics. A first one is especially designed to reduce the cost of navigating in the two-layer structure, in terms of \emph{rank}, \emph{select} and \emph{access} SDS operations. That is we try as much as possible to favor \emph{rank} operations against \emph{select} ones since most implementations guarantee constant time \emph{rank} operations on bitmap but not for \emph{select} ones which either need lot of extra space or logarithmic time. Two other heuristics are provided to take advantage of state of the art RDF access pattern \cite{DBLP:conf/www/StockerSBKR08,DBLP:conf/edbt/TsialiamanisSFCB12}, and statistics stored in the dictionary structures. 
Again, these heuristics have been adapted to reorder some access patterns which is a major source of optimizations for SPARQL queries containing many graph patterns. 
This results in the generation of query plans taking the form of left-deep join trees which is being translated and executed in terms of compositions of \emph{rank}, \emph{select} and \emph{access} SDS 
operations. In order to support \texttt{DISTINCT}, \texttt{LIMIT}, \texttt{OFFSET} and \texttt{ORDER BY} SPARQL operators, we provide a $k$-partite graph based storing system for the candidate tuples that allow us to store and filter them in an efficient way avoiding as much as possible unnecessary Cartesian product. Finally, the identifiers of the result are translated in terms of their associated values in the dictionaries. 

The supported SPARQL operators needed the development of optimization techniques in the query execution module: the \texttt{UNION} of graph patterns which is based on a lazy approach of common patterns, \texttt{FILTER} which requires accesses to the dictionary and \texttt{OPTIONAL} that prevents the creation of bindings in the absence of a matching for the optional graph patterns.

\section{Experimental evaluation}\label{experiments}
\subsection{System}
All experiments have been conducted on a HP Z800 workstation with 2 Quad-Core Intel Xeon Processors with 12Mbytes L2 cache, 8Gbytes of memory and running Gentoo 2.6.37 generic x86-64.
It contains two 500GB SATA disks running at 7200 rpm. We used gcc version 4.5.2 running on 64 bits with glibc 2.13.
We modified the libcds v1.0.13 in order to obtain \emph{rank\_prefix} and \emph{select\_prefix} operations on the proposed SDS.
We have compared our system with RDF-3X version 0.3.7, BigOWLIM version 3.5 and Jena 2.6.4 together with its TDB 0.8.10. 
We do not propose a comparison with Hexastore since it was not possible to load the data sets we are working with. 
This is due to its in-memory approach and the large number of set indexes, $i.e.,$ 6, it requires to process queries efficiently. 
Note that this aspect was confirmed in \cite{DBLP:conf/esws/Martinez-PrietoGF12} which essentially focuses on data loading, compression rates and times required for indexes creation.
Our current WaterFowl framework uses pointer-free wavelet trees (which were giving best results compared to pointer based wavelet trees and wavelet matrices).

\subsection{Data sets}
In this section, we present the results of our evaluation performed on a set of synthetic and a real world data sets.
The synthetic data sets correspond to instances of the Lehigh University Benchmark (LUBM) \cite{DBLP:journals/ws/GuoPH05}.
The main characteristics of LUBM are to feature an OWL ontology for the university domain, to enable scaling of data sets to an arbitrary size and to provide a set of 14 SPARQL queries of varying complexities.
Out of these queries, 10 require a form of inference, namely dealing with concept and property hierarchies as well as inverse and transitive roles which we are not testing since they require OWL entailments.
We are testing our system on two data sets, one for 100 and another one for 1000 universities.
Table \ref{data sets} summarizes the sizes in space and number of RDF triples of all data sets. The real world data sets corresponds to Yago and is mainly used on the first aspect of our evaluation.

\begin{table}
\centering
\caption{Description of the data sets}
\label{data sets}
\begin{tabular}{|c|r|r|} \hline
data set&Triples (Million)&Size (MB)\\ \hline
LUBM100 & 13.4 &1125\\ \hline
LUBM1000 & 133.5 & 11307\\ \hline
Yago2 & 37.5 &  5325 \\ \hline
\end{tabular}
\end{table}

\subsection{Results}
The results we are presenting in this section concern three aspects of our system: (i) memory footprint and time required to prepare a data set, (ii) query processing not requiring inferences and (iii) query answering requiring RDFS entailment regime.

The first one aims to demonstrate that a system designed on SDS possesses interesting properties in terms of data compression rate, time to prepare a data set, $i.e.,$ total duration required to create the dictionary, index the data, compute some statistics and serialize the database structure. 
It is presented in Table \ref{compress} and confirms the results contained in \cite{DBLP:conf/esws/Martinez-PrietoGF12}.
We can see that most compressed versions of WaterFowl, mode 2 and 3 relying respectively on non-pointer and so-called matrix wavelet trees require between 5 and 9\% of the space required by RDF-3X and this is even more important compared to BigOWLIM and Jena TDB. 
This is due to the high compression rate of the SDS we are using and the single, opposed to 15, index we are generating.
The sizes required for BigOWLIM and Jena TDB are explained by their approach which require full materialization.
Moreover, times to prepare a data set are about half of the duration taken by  RDF-3X. This is easily explained by the number of indexes RDF-3X is building.
Obviously, due to the materialization, the times needed to process and store the data sets are even more important for BigOWLIM and Jena TDB.
Finally, our ode 2, based on a wavelet tree pointer-free implementation seems to be an interesting trade-off between size of the generated data set and generation time.

\begin{table*}
\caption{Size of database serialization (MB) and Time to prepare data sets}
\center
\label{compress}
\begin{tabular}{|c|c|c|c|c|c|c|} \hline
& \multicolumn{3}{|c|}{Size in MB}& \multicolumn{3}{|c|}{Time in sec}\\ \cline{2-7}
& univ100 & univ1000 &Yago &univ100 & univ1000 & Yago\\ \hline
RDF-3X& 831,717& 7,795,458 & 2,189,735 & 240 & 3050& 1090 \\ \hline
BigOWLIM& 2,411,260 & 22,600,088 & 6,348,338 & 838 & 10640 & 3708 \\ \hline
Jena TDB & 1,492,057 & 13,984,467 & 3,928,271 & 1285 & 16332 & 5837 \\ \hline
WaterFowl Mode 1& 91,539 & 922,106 & 271,616 & 168 & 2134 &  768\\ \hline
WaterFowl Mode 2& 71,064 & 720396 & 210,556 & 119& 1515 & 545\\ \hline
WaterFowl Mode 3& 77,351 &798,829 & 203,728 & 107& 1488 &  513\\ \hline
\end{tabular}
\end{table*}

The two next aspects of our evaluation concerns query processing.
First, we consider queries that are not requiring reasoning. Then, we study some
queries requiring the RDFS entailment regime. 
We consider that by investigating both aspects of query answering, we are
able to highlight the pros and cons of our complete query processing
component.
Our evaluation methodology includes a warm-up phase before
measuring the execution time of the queries. This is required for the 3 compared systems but not for WaterFowl since its data reside in main-memory.
All the queries are first ran in sequence once to warm-up the systems,
and then the process is repeated 5 times.
The following tables report the mean values for each query and each system.

\begin{table*}
\caption{Query answering times (sec) on univ1000}
\center
\label{times}
\begin{tabular}{|c|c|c|c|} \hline 
& LUBM QR\#1 & LUBM QR\#2 &  LUBM QR\#14  \\ \hline
RDF-3X & 1.65  &  14.88 & 1640 \\ \hline
BigOWLIM & 138 & 5.7 & 3320\\ \hline
Jena TDB & 3.52 & 2.18 & 2998\\ \hline
WaterFowl Mode 2& 1.80 & 10.18 & 1710  \\ \hline
WaterFowl Mode 3& 1.75 &  10.13 & 1680 \\ \hline
\end{tabular}
\end{table*}

In the first context, we compare our approach with the 3 other systems on a subset of LUBM queries (\#1, \#2 and \#14).
Table \ref{times} emphasizes that the performances with the RDF-3X system are comparable.
Unsurprisingly, the two other systems are slower than RDF-3X on Queries \#1 and \#3.
A fact which has been highlighted on many other evaluations.
Note that these queries have different characteristics since they respectively correspond to large input with high selectivity,
complex 'triangle' query pattern and large input with low selectivity.
Query \#2 is performed more rapidly by Jena TDB and BigOWLIM but WaterFowl is faster than RDF-3X.
We consider that this is due to a better consideration of this query particular pattern.
It highlights that our query optimization has room for improvement.

\begin{table*}
\caption{Inference-based query answering times (sec) on univ100}
\center
\label{inferences}
\begin{tabular}{|c|c|c|c|c|c|} \hline 
& QR\#4 & QR\#5 & QR\#6 &  QR\#7 & QR\#10  \\ \hline
RDF-3X & 4.2 & 2.5 & 15.3  & 1.4 & 1.6  \\ \hline
OWLIM-SE & 705 & 16771 & 72 & 1708 & 3.65 \\ \hline
Jena TDB & 4.85 & 6.3 & 30.7 & 207 & 1.55 \\ \hline
WaterFowl Mode 2& 3.66 & 2.3 &13.4 & 1.2 & 1.4  \\ \hline
\end{tabular}
\end{table*}

In the context of queries requiring RDFS entailment, we are testing RDF-3X with
query rewriting performed using a DL reasoner against our system. 
That is, we have implemented a simple RDFS query rewriting on top of RDF-3X 
which generates SPARQL queries with \texttt{UNION} clauses. 
The RDF-3X approach enables to perform query rewriting in the context of the 
considered fastest RDF Store. 
Note that the two other systems do not require this machinery since they rely on a materialization
approach.
Table \ref{inferences} highlights that our system slightly outperforms the inference-enable RDF-3X on a set of five distinct LUBM queries, requiring different forms of reasoning, $i.e.,$ based on concept and property subsumption relationships.
It has already be emphasized that due to its large number of indices, RDF-3X is very competitive or 
even faster than materialization-based systems.
Due to our ontology elements encoding with prefix approaches and minimalist materialization
 of domain and range of properties, we outperform all systems on these five queries.

\section{Conclusion}
We have designed and implemented a novel type of RDF store that addresses a set of issues of big data and of the semantic web. Each database instance regroups a set of dictionaries and a data set represented in a compact, self-indexed manner using some succinct data structures.
The evaluation we have conducted emphasize that our system is clearly very efficient in terms of data compression and can thus be considered as an interesting alternative when one is concerned with data exchange. Moreover, on our query processing experimentations, our system presents performances that are comparable to the domain's reference, $i.e.,$ RDF-3X. We consider that this is quite a strong encouragement toward pursuing our work on WaterFowl.
 We consider that this is due to the advantage of our highly compressed data and implementing all data retrieving operations on SDS functions, $i.e.,$ \emph{access}, \emph{rank}, \emph{select} and their prefix counterparts. We also believe that founding and adapting all our query optimization heuristics on state of the art solutions is part of the good performances our system provides. Nevertheless, we are convinced that there is plenty of room for more optimizations in all modules of WaterFowl, $e.g.,$ pipelined parallelism in query execution.    

Our future experiments with WaterFowl suggest a promising direction for future investigations. They will mainly include the distribution of triples over a cluster of machines and the support for updates in both the TBox and the ABox of the knowledge base. Considering the latter, we are investigating incremental
solutions while an amortized approach is considered on the issue of updating the ABox. Finally, we would like to propose extensions of our ontology dictionary that go beyond the RDFS entailment regime, e.g. OWL2 entailment.

\bibliographystyle{abbrv}
\bibliography{waterfowl}  

\begin{thebibliography}{10}

\bibitem{DBLP:conf/vldb/AbadiMMH07}
D.~J. Abadi, A.~Marcus, S.~Madden, and K.~J. Hollenbach.
\newblock Scalable semantic web data management using vertical partitioning.
\newblock In {\em VLDB}, pages 411--422, 2007.

\bibitem{DBLP:conf/www/AtreCZH10}
M.~Atre, V.~Chaoji, M.~J. Zaki, and J.~A. Hendler.
\newblock Matrix "bit" loaded: a scalable lightweight join query processor for
  rdf data.
\newblock In {\em WWW}, pages 41--50, 2010.

\bibitem{DBLP:conf/semweb/BroekstraKH02}
J.~Broekstra, A.~Kampman, and F.~van Harmelen.
\newblock Sesame: A generic architecture for storing and querying rdf and rdf
  schema.
\newblock In {\em International Semantic Web Conference}, pages 54--68, 2002.

\bibitem{DBLP:conf/osdi/DeanG04}
J.~Dean and S.~Ghemawat.
\newblock Mapreduce: Simplified data processing on large clusters.
\newblock In {\em OSDI}, pages 137--150, 2004.

\bibitem{DBLP:conf/semweb/FernandezMG10}
J.~D. Fern{\'a}ndez, M.~A. Mart\'{\i}nez-Prieto, and C.~Gutierrez.
\newblock Compact representation of large rdf data sets for publishing and
  exchange.
\newblock In {\em International Semantic Web Conference (1)}, pages 193--208,
  2010.

\bibitem{DBLP:conf/kr/GottlobS12}
G.~Gottlob and T.~Schwentick.
\newblock Rewriting ontological queries into small nonrecursive datalog
  programs.
\newblock In {\em KR}, 2012.

\bibitem{DBLP:conf/soda/GrossiGV03}
R.~Grossi, A.~Gupta, and J.~S. Vitter.
\newblock High-order entropy-compressed text indexes.
\newblock In {\em SODA}, pages 841--850, 2003.

\bibitem{DBLP:conf/pods/GrossiO12}
R.~Grossi and G.~Ottaviano.
\newblock The wavelet trie: maintaining an indexed sequence of strings in
  compressed space.
\newblock In {\em PODS}, pages 203--214, 2012.

\bibitem{DBLP:journals/ws/GuoPH05}
Y.~Guo, Z.~Pan, and J.~Heflin.
\newblock Lubm: A benchmark for owl knowledge base systems.
\newblock {\em J. Web Sem.}, 3(2-3):158--182, 2005.

\bibitem{harris_sparql11_2013}
S.~Harris and A.~Seaborne.
\newblock {SPARQL} 1.1 query language {W3C} recommendation.
  http://www.w3.org/tr/sparql11-query/, 2013.

\bibitem{hayes_rdf_2004}
P.~Hayes.
\newblock {RDF} semantics, {W3C} recommendation. http://www.w3.org/tr/rdf-mt/,
  2004.

\bibitem{DBLP:conf/focs/Jacobson89}
G.~Jacobson.
\newblock Space-efficient static trees and graphs.
\newblock In {\em FOCS}, pages 549--554, 1989.

\bibitem{DBLP:conf/esws/Martinez-PrietoGF12}
M.~A. Mart\'{\i}nez-Prieto, M.~A. Gallego, and J.~D. Fern{\'a}ndez.
\newblock Exchange and consumption of huge rdf data.
\newblock In {\em ESWC}, pages 437--452, 2012.

\bibitem{DBLP:conf/www/MotikHS07}
B.~Motik, I.~Horrocks, and U.~Sattler.
\newblock Bridging the gap between owl and relational databases.
\newblock In {\em WWW}, pages 807--816, 2007.

\bibitem{DBLP:conf/fsttcs/Munro96}
J.~I. Munro.
\newblock Tables.
\newblock In {\em FSTTCS}, pages 37--42, 1996.

\bibitem{DBLP:conf/sigmod/NeumannW09}
T.~Neumann and G.~Weikum.
\newblock Scalable join processing on very large rdf graphs.
\newblock In {\em SIGMOD Conference}, pages 627--640, 2009.

\bibitem{DBLP:journals/vldb/NeumannW10}
T.~Neumann and G.~Weikum.
\newblock The rdf-3x engine for scalable management of rdf data.
\newblock {\em VLDB J.}, 19(1):91--113, 2010.

\bibitem{DBLP:conf/semweb/Perez-UrbinaHM09}
H.~P{\'e}rez-Urbina, I.~Horrocks, and B.~Motik.
\newblock Efficient query answering for owl 2.
\newblock In {\em International Semantic Web Conference}, pages 489--504, 2009.

\bibitem{revuz91}
D.~Revuz.
\newblock {\em Dictionnaires et lexiques : methodes et algorithmes}.
\newblock PhD thesis, Paris 7, 1991.

\bibitem{DBLP:conf/kr/Rodriguez-MuroC12}
M.~Rodriguez-Muro and D.~Calvanese.
\newblock High performance query answering over dl-lite ontologies.
\newblock In {\em KR}, 2012.

\bibitem{DBLP:conf/kr/RosatiA10}
R.~Rosati and A.~Almatelli.
\newblock Improving query answering over dl-lite ontologies.
\newblock In {\em KR}, 2010.

\bibitem{DBLP:conf/www/StockerSBKR08}
M.~Stocker, A.~Seaborne, A.~Bernstein, C.~Kiefer, and D.~Reynolds.
\newblock Sparql basic graph pattern optimization using selectivity estimation.
\newblock In {\em WWW}, pages 595--604, 2008.

\bibitem{DBLP:conf/edbt/TsialiamanisSFCB12}
P.~Tsialiamanis, L.~Sidirourgos, I.~Fundulaki, V.~Christophides, and P.~A.
  Boncz.
\newblock Heuristics-based query optimisation for sparql.
\newblock In {\em EDBT}, pages 324--335, 2012.

\bibitem{DBLP:journals/pvldb/WeissKB08}
C.~Weiss, P.~Karras, and A.~Bernstein.
\newblock Hexastore: sextuple indexing for semantic web data management.
\newblock {\em PVLDB}, 1(1):1008--1019, 2008.

\end{thebibliography}
\section{Appendix}
For completeness, we include the SPARQL queries used in our evaluation.
We also include the encodings of Query 1 (requiring no inference) and of Query 4 (requiring inference).

The following prefixes are used all along the queries:\\
rdf: <http://www.w3.org/1999/02/22-rdf-syntax-ns\#>\\
lubm: <http://www.lehigh.edu/~zhp2/2004/0401/\\univ-bench.owl>

\textbf{Query 1}\\
SELECT ?x WHERE \{ \\
?x rdf:type lubm:GraduateStudent. \\
?x lubm:takesCourse\\
<http://www.Department0.University0.edu/\\GraduateCourse0>.\}\\

Considering that the encodings of the concept GraduateStudent is 01100, of the takesCourse object property is 1001000 
and of http://www.Department0.University0.edu/\\GraduateCourse0 is 1000010101, the BGPs of this query are encoded as follows:\\
\{ \\
?x 00 0100.\\
?x 1001000 1000010101.\}

\textbf{Query 2}\\
SELECT ?x ?y ?z WHERE \{\\ 
	?x rdf:type lubm:GraduateStudent . \\
	?y rdf:type lubm:University . \\
	?z rdf:type lubm:Department . \\
	?x lubm:memberOf ?z . \\
	?z lubm:subOrganizationOf ?y .\\ 
	?x lubm:undergraduateDegreeFrom ?y.\}\\
	
\textbf{Query 4}\\
SELECT ?x ?y1 ?y2 ?y3 WHERE \{\\
	?x rdf:type lubm:Professor . \\
	?x lubm:worksFor \\
	<http://www.Department0.University0.edu>. \\
	?x lubm:name ?y1 . \\
	?x lubm:mailAddress ?y2 .\\
	?x lubm:telephone ?y3.\}\\	

For this encoding, all ontology identifiers are given in Figure \ref{encodings} and the URI http://www.Department0.University0.edu is encoded as 1001001
	
\{\\
?x 00 010101011 . \\
?x 001111 1001001. \\
?x 01010 ?y1 . \\
?x 01001 ?y2 .\\
?x 01101 ?y3.\}\\

\textbf{Query 5}\\
SELECT ?x WHERE \{ \\
	?x rdf:type lubm:Person . ?x lubm:memberOf\\
	 <http://www.Department0.University0.edu>.\}

\textbf{Query 6}\\
SELECT ?x WHERE \{ \\
	?x rdf:typetype lubm:Student.\}

\textbf{Query 7}\\
SELECT ?x ?y WHERE \{ \\
	?x rdf:type lubm:Student . \\
	?y rdf:type lubm:Course . \\
	<http://www.Department0.University0.edu/\\AssociateProfessor0>\\
	lubm:teacherOf ?y .\\ 
	?x lubm:takesCourse ?y.\}	
	
\textbf{Query 10}\\
SELECT ?x WHERE \{ \\
	?x rdf:type lubm:Student . ?x lubm:takesCourse \\
	<http://www.Department0.University0.edu/\\GraduateCourse0>.\}

\textbf{Query 14}\\
SELECT ?x WHERE \{ \\
	?x rdf:type lubm:UndergraduateStudent .\}\\	
\end{document}